\documentclass[twoside]{article}
\usepackage{fleqn,espcrc2}
\usepackage{graphicx}
\usepackage[figuresright]{rotating}
\usepackage{epsfig}
\newcommand{\vgas}{\ensuremath{\mathrm V_{gas}}}
\newcommand{\qgamma}{\ensuremath{\mathrm Q_\gamma}}

\title{First results from an aging test of a prototype RPC for the
  LHCb Muon System}

\author{A. Bizzeti\address[FIRE]{University of Florence and
  INFN -- Florence, Via G. Sansone 1, I--50019, Sesto F.no, Florence,
  Italy}, G. Carboni\address[ROM2]{University of Rome ``Tor Vergata''
  and INFN -- Rome II, Via della Ricerca Scientifica 1, I--00133 Rome
  Italy}, G. Collazuol\addressmark[FIRE], S. De
  Capua\addressmark[ROM2], D. Domenici\addressmark[ROM2],
  G. Ganis\addressmark[ROM2], R. Messi\addressmark[ROM2],\\ G. Passaleva\addressmark[FIRE]\thanks{corresponding author}, E. Santovetti\addressmark[ROM2], M. Veltri\addressmark[FIRE] %
}%
      
\begin{document}

\begin{abstract}
Recent results of an aging test performed at the
CERN Gamma Irradiation Facility on a single--gap RPC prototype
developed for the LHCb Muon System are presented. The results are based on an
accumulated charge of about 0.45 C/cm$^2$, corresponding to about 4
years of LHCb running at the highest background rate. The performance
of the chamber has been studied under several photon flux values
exploiting a muon beam. A degradation of the rate capability
above 1 kHz/cm$^2$ is observed, which can be correlated to a
sizeable increase of resistivity of the chamber plates. An
increase of the chamber dark current is also observed. 
The chamber performance is found to fulfill the LHCb operation requirements.
\vspace{1pc}
\end{abstract}

\maketitle

\section{Introduction}

Resistive Plate Chamber (RPC) detectors have been adopted by the LHCb
Collaboration to cover a large fraction (about
48\%) of the Muon System~\cite{TDR,TDR_RPC}. The LHCb
experiment~\cite{TP} covers the forward part of the solid angle and
is therefore subject to a very large particle flux. 
In particular, the particle rates expected in the
Muon System are significantly larger than those expected by
ATLAS~\cite{atlas_tdr} and CMS experiments~\cite{cms_tdr}.
In the regions
covered by RPCs, the maximum particle rate is expected to vary
between 0.25 and 0.75 kHz/cm$^2$,
depending mainly on the polar angle. These large rates are potentially
dangerous from the point of view of the RPC aging, which must be
carefully studied and evaluated.

The main aging effect for RPCs is produced by the current flowing
through the resistive plates; It has been demonstrated, indeed, that the
irradiation of bakelite slabs with photons up to an integrated dose of
20 kGy does not produce any degradation in the bakelite
properties~\cite{danilo}. The most important parameter to
evaluate the aging effects is given therefore by the total charge flow
accross the detector during its lifetime . Assuming
an average avalanche charge of 30 pC~\cite{danilo}, the total charge
integrated by LHCb RPCs over 10 years of operation range
from 0.35 C/cm$^2$, in the regions of the Muon System where the particle flux
is smaller, to about 1.1 C/cm$^2$ in the high rate regions.

Aging tests performed by the RPC groups of ATLAS and CMS have shown
that their detectors can withstand the radiation doses expected in
those experiments. However, these tests are based on an integrated
charge of at most 0.3 C/cm$^2$~\cite{ajello,gabbriello} which is four
times smaller than the maximum  expected in LHCb RPCs.

To cope with the LHCb requirements, an extensive aging programme has
been devised, which started in January 2001
and is expected to last, at a first stage, until December 2001,
exploiting the large CERN Gamma Irradiation Facility (GIF)~\cite{GIF}, where a
$^{137}$Cs gamma source of about 655 GBq is available.

\section{\boldmath Setup of the aging test}

The aging test consists in irradiating for about one year a single gap RPC
({\it irradiated RPC})  at the GIF facility. The position 
closest to the source available for the chamber is at about 1 meter,
where the photon flux is, roughly, 1.5 kHz/cm$^2$. Another chamber ({\it
  reference RPC}) is installed just outside the irradiation area.
Both chambers share the same gas and high voltage lines so that
they are operated in the same conditions. In this way it
is possible 
to disentangle the effects due to variations in the environment
parameters, such as temperature, pressure or gas quality, from those due
to the irradiation.

\begin{figure}[!ht]
 \includegraphics[width=7.2cm]{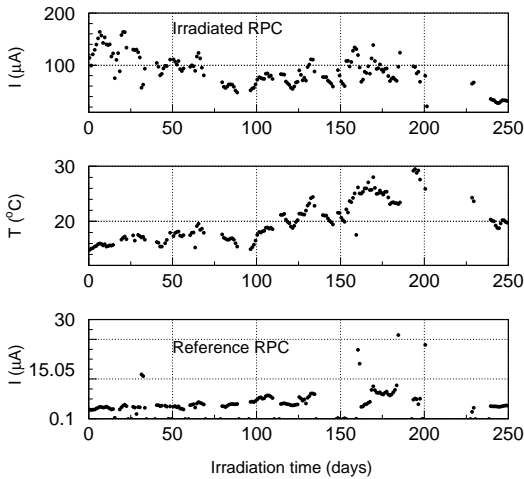}
 \caption{Current drawn by the irradiated RPC as a function of
   irradiation time (top), monitored temperature (middle) and
   current drawn by the
   reference chamber (bottom).}
 \label{fig:current_irr} 
\end{figure}

The two RPCs prepared for the aging test have resistive plates made
of phenolic bakelite with resistivity $\rho = 9\cdot 10^9\
\Omega$~cm. The plates are oiled with linseed oil. The size of both
chambers is 50x50 cm$^2$.
The readout strips of the irradiated RPC are 3 cm wide and have been cut 
in two along their length, so that the final strip area is 3x25 cm$^2$. 
The strips of the reference chamber, instead, are 3 cm wide and 50 cm long. 
The RPCs are operated with a gas mixture consisting of 95\% $\rm
C_2H_2F_4$, 4\% $\rm iC_4H_{10}$, 1\% $\rm SF_6$.
To simulate the LHCb conditions and perform the test in a reasonable
amount of time, the applied voltage was chosen in such a way to
have an average
avalanche charge of about 50 pC, that, at the rate quoted above,
yields a current density of about 80 nA/cm$^2$. This, with a realistic
duty factor of 30\%, gives an integrated
charge of about 0.8 C/cm$^2$ in one year of irradiation, which corresponds to
about 7 LHCb years in the worst background conditions. 

\begin{figure}[!ht]
 \includegraphics[width=7.2cm]{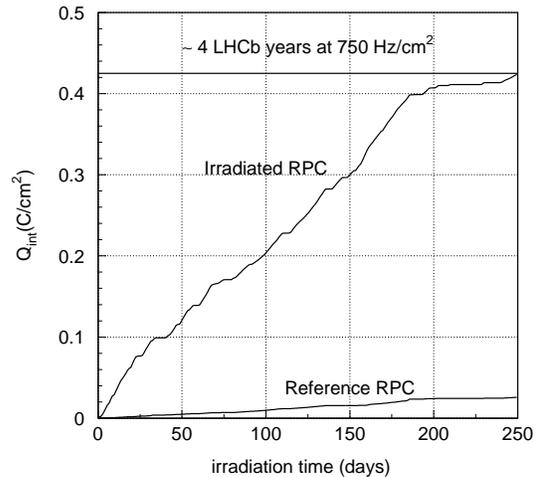}
 \caption{Integrated currents versus time from the beginning of the
   aging test}
 \label{fig:current_int} 
\end{figure}

All the relevant parameters of the test,
temperature, pressure, high voltage and currents drawn by the chambers,
are continuously monitored and recorded on a PC. In
Figure~\ref{fig:current_irr} the current drawn by the irradiated RPC is
shown as a function of time (top plot) along with the temperature
(middle plot). In
the bottom plot the current drawn
by the reference RPC is also shown. The latter is 
reasonably
stable and constant over the whole aging test, showing that
the chambers have been operated rather smoothly and no large systematic
effects are expected to affect the performance of the two RPCs.
In Figure~\ref{fig:current_int}, the integrated
charge as a function of time is shown. The irradiated RPC has
accumulated about 0.45 C/cm$^2$ corresponding to about 4 LHCb years
in the regions with the highest background rate and to more than 10
LHCb years in the low background regions.

To check the performance of the irradiated chamber at this
intermediate aging stage, a beam test was performed, using the
muon beam available at the GIF.
In the next sections the results of this
test will be described.

\begin{figure}[!ht]
 \includegraphics[width=7.2cm]{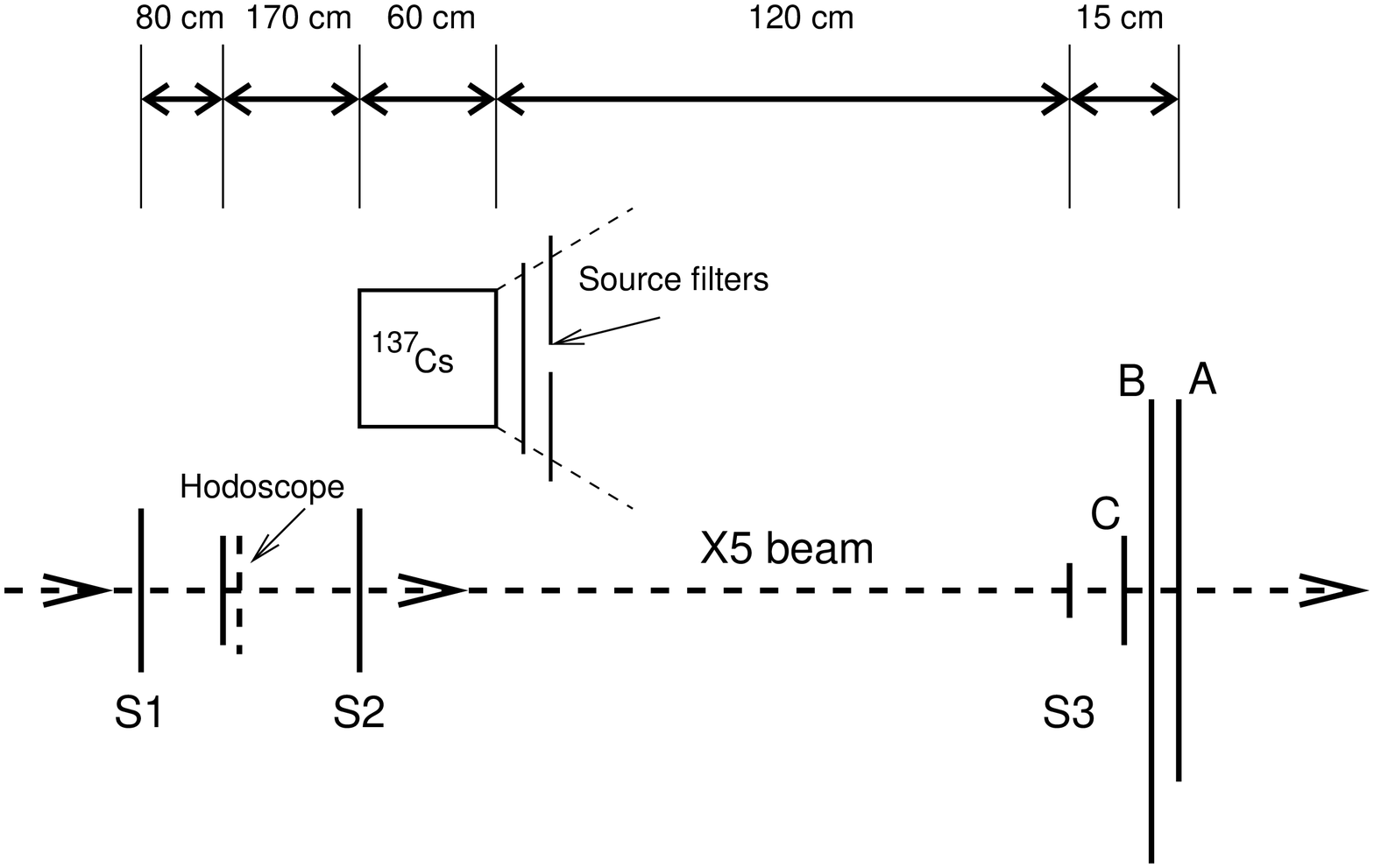}
 \caption{Setup of the test at the GIF. A=irradiated RPC;
   B=reference RPC; C=10$\times$10 cm$^2$ RPC. }
 \label{fig:test_setup}
\end{figure}

\section{\boldmath Beam test results}

The irradiated and the reference RPCs have been carefully tested
exploiting the muon beam available at the GIF. The test setup is
sketched in Figure~\ref{fig:test_setup}. The combined use of the
gamma source and the beam allows a
test of the chamber performance parameters, like efficiency, cluster
size and rate capability, under different photon fluxes, ranging from zero
(source off) up to a maximum that, in this test setup, was about 1
kHz/cm$^2$. The chambers were operated in the same conditions as in the aging
test from the point of view of the gas and high voltage supplies.
The chamber efficiency was evaluated by tracking the beam particles
with a scintillator hodoscope and with an additional small $10\times
10$ cm$^2$ RPC. The photon fluxes were estimated from the chamber
counting rates measured during dedicated off-spill gates.
The measurements described in the next sections were obtained at three
different source attenuation values, namely attenuation 1, 2 and 5
corresponding roughly to 1 kHz/cm$^2$, 0.7 kHz/cm$^2$ and 0.4
kHz/cm$^2$ photon fluxes respectively.

\subsection{\boldmath Analysis of currents}

\begin{figure}[!ht]
 \includegraphics[width=7.2cm]{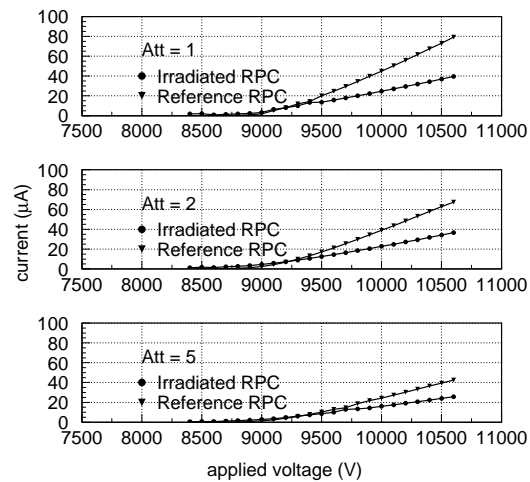}
 \caption{RPC currents versus the applied voltage at different
   photon fluxes}
 \label{fig:currents}
\end{figure}

The current drawn by the irradiated chamber with the source on, was
found to be about
a factor of two smaller than that of the reference RPC as can be seen
from Figure~\ref{fig:currents}.

\begin{figure}[!ht]
 \includegraphics[width=7.2cm]{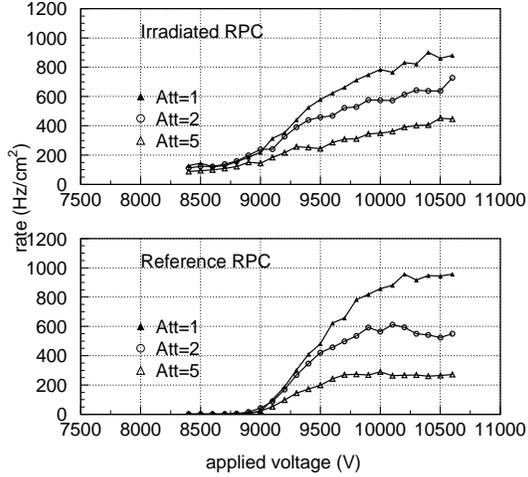}
 \caption{RPC counting rates as a function of the applied voltage}
 \label{fig:rates}
\end{figure}

\begin{figure}[!ht]
 \includegraphics[width=7.2cm]{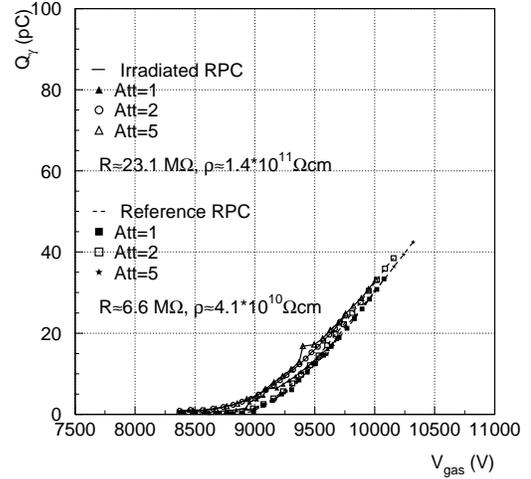}
 \caption{Distribution of \qgamma\ as a function of \vgas\ for the two
   RPCs at different photon fluxes}
 \label{fig:qgamma}
\end{figure}

This effect can be interpreted within a model introduced in
~\cite{aielli1}. In this simple model one assumes
that under a high particle flux, the RPC working point is
determined by an effective voltage $\rm \vgas = V - IR $
where V is the applied voltage, I is the current drawn by the chamber
and R is the resistance of the resistive plates. In
this framework, quantities such as the average avalanche charge
\qgamma\ or the
chamber efficiency $\varepsilon$ are only functions of the gas
properties and of \vgas. This means that once the correct value of the
resistance R has been determined, the functions $\varepsilon(\vgas)$ and
$\qgamma(\vgas)$ must be universal functions independent on the
particle flux and on the plate characteristics. This method
allows a non destructive estimation of the plate
resistivity, that can be monitored on--line during the aging test.

\qgamma{} is measured by normalising the current drawn by the chambers
to the rate of converted photons, which can be obtained from the
chamber counting rate at the plateau, corrected for the efficiency.
The measured counting rates are plotted in Figure~\ref{fig:rates}. 

The
distributions of \qgamma\ for the two RPCs, at different
photon fluxes, are plotted as a function of \vgas\ in
Figure~\ref{fig:qgamma}; they are fairly consistent with a single
curve. The resulting values of the plate resistance, for the
irradiated and the reference RPCs are $\rm R_{irr} = 23.1 M\Omega$ and
$\rm R_{ref} = 6.6 M\Omega$ respectively corresponding to equivalent
bakelite resistivities $\rm \rho_{irr} = 1.4\times 10^{11} \Omega cm$
and $\rm \rho_{ref} =  4.1\times 10^{10} \Omega cm$.
Compared to the original resistivity value and to the reference
chamber, a large
increase in the plate resistivity for the irradiated RPC can be
observed, 
which can
be ascribed to the radiation effect and that accounts for the current
drop mentioned above.

\begin{figure}[htb]
 \includegraphics[width=7.2cm]{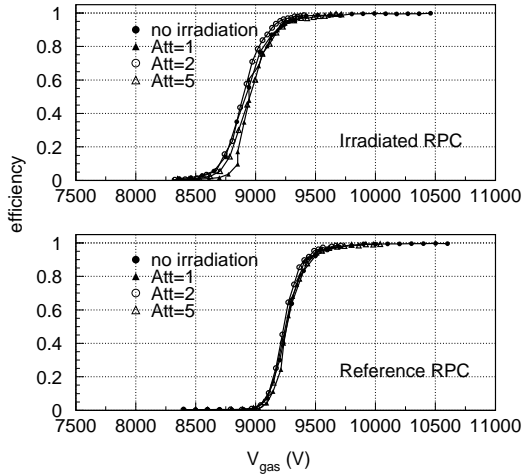}
 \caption{Efficiencies versus \vgas{} at different photon
   fluxes.}
 \label{fig:efficiencies}
\end{figure}

The efficiencies of the two RPCs are plotted as a function of \vgas{}
in Figure~\ref{fig:efficiencies}. Again, the curves are consistent
with an universal function $\varepsilon(\vgas)$.

\subsection{\boldmath Efficiencies and rate capabilities}

\begin{figure}[htb]
 \includegraphics[width=7.2cm]{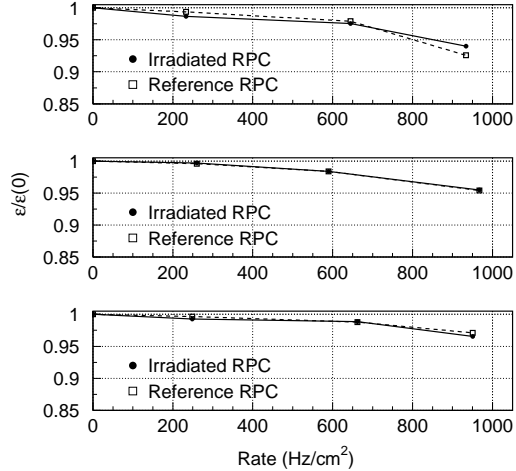}
 \caption{Efficiency versus rate for the two RPCs at three different
   high voltage values. The efficiencies are normalised to those
   obtained with the source off}
 \label{fig:rate_capability}
\end{figure}

Because of the stringent requirements imposed by the LHCb muon
trigger~\cite{TDR}, 
the RPCs are required to have an efficiency above 95\% up to
the maximum expected background rate.
In order to test the ability of the irradiated RPC to work within the
LHCb operational parameters, the rate capability of the two RPCs has
been carefully tested.
In
Figure~\ref{fig:rate_capability} the RPC efficiency as a function of
the photon flux for three different high voltage values is
shown. The efficiency values are normalised to those obtained with the
source off. At the nominal applied voltage of 10.2 kV, 
the efficiency exceeds 95\% for both chambers up to a photon flux of about 0.8
kHz/cm$^2$. However, as soon as the high voltage is raised by a few
hundreds volts, the RPCs recover their full efficiency up to a flux of
more than 1 kHz/cm$^2$. Moreover there is no evidence of a different
behaviour between the irradiated and the reference RPC. 
From this test
it is possible to conclude that the RPCs under test can be safely
operated, from the aging point of view, at least for 4 LHCb
years in the worst expected background conditions.

\subsection{\boldmath Dark currents}

In the design of RPCs for LHCb, the level of dark currents drawn by
the chambers is of concern, especially from the aging point of view.
The dark currents will contribute an aging effect that must be kept
well below the aging caused by the particle flux.
Considering as acceptable an extra-aging of 25\%, 
the dark current density must be kept below 3 nA/cm$^2$, over the
experiment lifetime.

\begin{figure}[htb]
 \includegraphics[width=7.2cm]{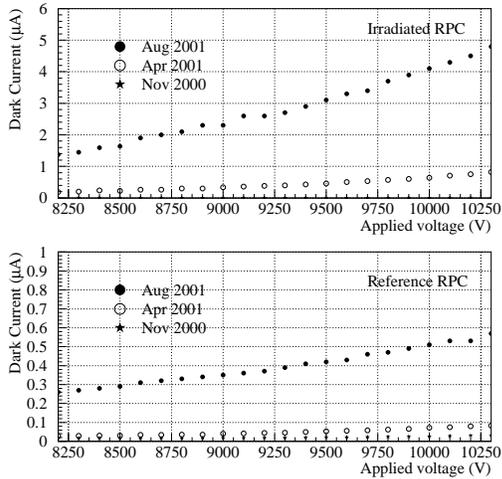}
 \caption{Dark currents, measured at different intermediated times of
   the aging test, versus the applied voltage.}
 \label{fig:dark_currents}
\end{figure}

To
check if these constraints are fulfilled,
the behaviour of the dark currents of the chambers under test has
been studied. In
Figure~\ref{fig:dark_currents} the dark currents drawn by the two RPCs
as a function of the applied voltage, measured at different
intermediate times during the aging test are shown. A large increase
of the dark current drawn by the irradiated RPC can be clearly
observed. The effect of the irradiation amounts to about 2 nA/cm$^2$. This
behaviour is still under investigation and
no interpretation can be worked out yet. Anyhow, it is worth to stress
that the dark current observed in the irradiated chamber is
still below the maximum allowed value for the LHCb Muon System.

\section{Summary and conclusions}

The results of an aging test of an RPC prototype for
the LHCb muon system have been described.
The irradiated prototype has been tested with a muon beam at the GIF
after a charge of 0.45 C/cm$^2$ has been integrated, corresponding to
about 4 years of LHCb at the highest expected background rate and
to more than 10 LHCb years in the lower rate regions of the Muon
System. The performance of the irradiated RPC has been compared to
that of a reference chamber. As an effect of the
irradiation, a large
increase of the resistivity of the chamber plates in the irradiated
RPC has been observed; a clear increase of the dark current is also observed.
The irradiated RPC prototype is still well efficient up to the maximum rate
expected in the LHCb muon system and the dark current level is under control.

\end{document}